# Graphene quantum dots for valley-based quantum computing:

# A feasibility study

**G. Y. Wu[1,2]\*, N.-Y. Lue[2], and L. Chang[2]**

[1]Department of Electrical Engineering, National Tsing-Hua University, Hsin-Chu 30013, Taiwan, ROC;

[2]Department of Physics, National Tsing-Hua University, Hsin-Chu 30013, Taiwan, ROC; \*e-mail: yswu@ee.nthu.edu.tw

At the center of quantum computing[1] realization is the physical implementation of qubits – two-state quantum information units. The rise of graphene[2] has opened a new door to the implementation. Because graphene electrons simulate two-dimensional relativistic particles with two degenerate and independent energy valleys,[3] a novel degree of freedom (d.o.f.), namely, the valley state of an electron, emerges as a new information carrier.[4] Here, we expand the Loss-DiVincenzo quantum dot (QD) approach in electron spin qubits,[5,6] and investigate the feasibility of double QD (DQD) structures in gapful graphene as "valley qubits", with the logic 0 / 1 states represented by the "valley" singlet / triplet pair. This generalization is characterized by 1) valley relaxation time ~ O(ms), and 2) electric qubit manipulation on the time scale ~ ns, based on the 1st-order "relativistic effect" unique in graphene. A potential for valley-based quantum computing is present. (arXiv 1104.0443 cond-mat.mes-hall Apr. 4, 2011)

**(The framework)** A qubit implementation faces three important issues, namely, i) all-electrical manipulation, ii) state relaxation / decoherence, and iii) scalability and fault-tolerance. The recent advance in electron spin qubit research provides, for the development of analogous qubits, a framework to address these issues. In the spin case, the paradigm QD approach (– using confined electron spins)[5,6] usually serves as the foundation, upon which one applies the additional tactics including: utilization of the Rashba mechanism of spin-orbit interaction (SOI) to achieve i),[7-9] materials with weak SOI and vanishing hyperfine field (HF), e.g., graphene[10] or carbon nanotube (CNT)[11], to resolve ii), and spin singlet-triplet qubits to iii).[12-15]

Being solutions to separate issues, these tactics are sometimes at odds with one another, in a material-dependent way. For instance, in materials with strong Rashba SOI, HF or SOI inevitably cause state mixing.[16-18] For this reason, varied materials, e.g., GaAs,[19-21] CNT[11], or InAs,[22] have been exploited, in the recent experimental breakthroughs in spin qubit demonstration.

The rising graphene opens a new path to approach the issues. Here, we investigate valley singlet-triplet qubits in graphene DQDs.[23] Notably, in this proposal, the aforesaid tactics developed for spin qubits are all realized (in their "valley" version) without contradicting one another.

**(Quantum-mechanical description)** The QDs envisioned here are electrostatically defined in a gapful graphene epitaxial layer (grown on a SiC or BN substrate). A free electron in the layer obeys the Dirac-type equation,[3] with energy dispersion characterized by a finite band gap ($2\Delta$),[24,25] an effective mass $m^* = \Delta/v_F^2$ ($v_F$ = *Fermi velocity*), as well as two degenerate and inequivalent energy valleys (denoted as K and K', to which we will attach indices $\tau_v = \pm 1$ below). We consider near-gap, QD-confined electrons, and study their interaction with external electric and magnetic fields for the manipulation of valley d.o.f.. The Dirac equation is replaced by the following "non-relativistic Schrodinger equation" (including the 1st-order "relativistic correction" (R.C.)),[26]

$$H(\tau_v)\phi \approx E(\tau_v)\phi, \tag{1}$$
$$H(\tau_v) = H^{(0)}(\tau_v) + H^{(1)}(\tau_v).$$

$H^{(0)}$ is the non-relativistic part, with

$$H^{(0)} = \frac{\vec{\pi}^2}{2m^*} + V + \tau_v \mu_{v0} B_{normal} \quad (\mu_{v0} \equiv \frac{e\hbar}{2m^*}).$$

$H^{(1)}$ is the 1st-order R.C., with

$$H^{(1)} = -\frac{1}{2\Delta}\left(\frac{\vec{\pi}^2}{2m^*} + \tau_v \mu_{v0} B_{normal}\right)^2 + \tau_v \frac{\hbar}{4m^*\Delta}(\nabla V) \times \vec{\pi} - \frac{1}{8m^*\Delta}(\vec{p}^2 V).$$

Here,

$$V = V_{QD} + V_\varepsilon, \quad \vec{\pi} \equiv \vec{p} + e\vec{A}.$$

$V_{QD}$ is the QD confinement potential energy, $V_\varepsilon$ the potential energy in the electric field, **B**$_{normal}$ the magnetic field (normal to the epi-layer), and **A** the corresponding vector potential.

The $\tau_v$-dependent terms in $H^{(0)}$ and $H^{(1)}$ represent the coupling of valley d.o.f. to external fields. They give rise to the following property useful for valley manipulation.

**(Valley magnetic moment, $\mu_v$, and tuning)** For $\mathbf{B_{normal}} \neq 0$, the valley degeneracy is lifted.[27,28] The last term in $H^{(0)}$ gives the non-relativistic "Zeeman splitting", $E_Z = \pm\mu_{v0} B_{normal}$, independent of the electron energy. This result may be extended to include the 1st-order R.C.. We consider a QD where $V_{QD} = \frac{1}{2} m^* w_0^2 r^2 + \frac{1}{4} k_{4x} m^* w_0^2 x^4$. (For $k_{4x} < 0$, it describes a parabolic potential with a cut-off in the x-direction.) For $k_{4x} \ll m^* w_0/\hbar$, a perturbative calculation using Eqn. (1) yields $E_Z = \pm\mu_v B_{normal}$, with the magnetic moment[26]

$$\mu_v \approx \mu_{v0}\left(1 - \frac{\hbar w_0}{4\Delta}\right), \quad \text{(for } \hbar w_0 \ll \Delta, \ k_{4x} \ll \frac{m^* w_0}{\hbar}\text{)} \tag{2}$$

$\hbar w_0$ here is the ground state energy. Notably, with the R.C., $\mu_v$ is now energy-dependent. Thereby, a DC electric field, e.g., $\varepsilon_x^{(DC)}$ (in the x-direction), can be applied to shift the electron energy and achieve $\mu_v$-tuning, with[26]

$$\delta\mu_v^{(DC)} \approx -\frac{3}{16}\mu_{v0}\left(k_{4x} x_\varepsilon^{(DC)2}\right)\frac{\hbar w_0}{\Delta} \quad \text{(for } \varepsilon_x^{(DC)} \ll \sqrt{\hbar m^* w_0^3}/e\text{)} \tag{3}$$

$$(x_\varepsilon^{(DC)} \equiv -\frac{e\varepsilon_x^{(DC)}}{m^* w_0^2})$$

being the field-induced change in $\mu_v$. ($x_\varepsilon^{(DC)}$ = *field-induced electron displacement*.)

We note that $\delta\mu_v^{(DC)}$, being a 1st-order R.C., derives from the two valley-dependent terms in $H^{(1)}$. One of them is "$\tau_v(\hbar/4m^*\Delta)\nabla V \times \boldsymbol{\pi}$", the valley-orbit interaction (VOI)[29] and an analogue of SOI. Being valley-diagonal, the VOI (as well as the other term) is free of state-flipping/state-mixing effects, as opposed to the SOI.

AC mode of $\mu_v$-tuning can also be achieved, when an AC field is superimposed on the DC field, e.g., $\varepsilon_x = \varepsilon_x^{(DC)} + \varepsilon_x^{(AC)} \sin(w_{AC}t)$. Under the adiabatic condition, e.g., $w_{AC} \ll w_0$, where the AC field does not cause any transition between QD energy levels, the following additional AC tuning ($\delta\mu_v^{(AC)}$) is obtained, with[26]

$$\delta\mu_v = \delta\mu_v^{(DC)} + \delta\mu_v^{(AC)},$$

$$\delta\mu_v^{(AC)} \approx -\frac{3}{8}\mu_{v0}\left(k_{4x}x_\varepsilon^{(DC)}x_\varepsilon^{(AC)}\right)\sin(w_{AC}t)\frac{\hbar w_0}{\Delta}, \quad \text{(for } \varepsilon_x^{(AC)} \ll \varepsilon_x^{(DC)}) \qquad (4)$$

$$x_\varepsilon^{(AC)} \equiv -\frac{e\varepsilon_x^{(AC)}}{m^* w_0^2}.$$

**(Qubit states)** Fig. 1(a) shows the DQD qubit structure, with one electron residing in each QD. In the absence of magnetic field, each QD is assumed to accommodate only one bound state (with both spin and valley degeneracy). An in-plane magnetic field, **B**$_{plane}$, which couples to the spin but not the valley d.o.f., is used to lift the spin degeneracy. See Fig. 1(b). $\{K_L, K_L'\}$ / $\{K_R, K_R'\}$ denote the lower valley doublets of the spin-split states in the left / right QDs, respectively. Between the electrons, there is the exchange interaction

$$H_J = \frac{1}{4}J\vec{\tau}_L \cdot \vec{\tau}_R, \qquad (5)$$

where $\tau_{L(R)}$ = "*Pauli valley operator*", with

$$\tau_x = \begin{pmatrix} 0 & 1 \\ 1 & 0 \end{pmatrix}, \tau_y = \begin{pmatrix} 0 & -i \\ i & 0 \end{pmatrix}, \tau_z = \begin{pmatrix} 1 & 0 \\ 0 & -1 \end{pmatrix}.$$

J = *exchange integral* ~ $4t_{d-d}^2/U$, $t_{d-d}$ = *interdot tunneling*, and U = *on-site Coulomb energy*.

Generally, J = O(meV) is achievable in the QD approach.[10,26]

The logic 0 / 1 states are represented by the "valley" singlet $|z_S\rangle$ / triplet $|z_{T0}\rangle$. Here,

$$|z_S\rangle = \frac{1}{\sqrt{2}}(c_{K_L}^+ c_{K_R'}^+ - c_{K_L'}^+ c_{K_R}^+)|vaccum\rangle, \qquad (6)$$

$$|z_{T0}\rangle = \frac{1}{\sqrt{2}}(c_{K_L}^+ c_{K_R'}^+ + c_{K_L'}^+ c_{K_R}^+)|vaccum\rangle,$$

in terms of electron creation operators. Linear combinations of $\{|z_S\rangle, |z_{T0}\rangle\}$ give

$$|x_+\rangle = c_{K_L}^+ c_{K_R'}^+|vaccum\rangle, \quad |x_-\rangle = c_{K_L'}^+ c_{K_R}^+|vaccum\rangle.$$

The qubit state space (denoted as $\mathbf{\Gamma_v}$) is expanded by $|z_S\rangle$ and $|z_{T0}\rangle$, and isomorphic to the spin-1/2 state space, e.g., $|z_S\rangle \leftrightarrow |s_z = -1/2\rangle$, $|z_{T0}\rangle \leftrightarrow |s_z = 1/2\rangle$, $|x_-\rangle \leftrightarrow |s_x = -1/2\rangle$, $|x_+\rangle \leftrightarrow |s_x = 1/2\rangle$. The other triplet states,

$$|z_{T+}\rangle = c_{K_L}^+ c_{K_R}^+|vaccum\rangle, \quad |z_{T-}\rangle = c_{K_L'}^+ c_{K_R'}^+|vaccum\rangle,$$

are outside $\mathbf{\Gamma_v}$ and not needed for quantum computing. $\mathbf{B_{normal}}$ splits $|z_{T\pm}\rangle$ away from $\{|z_S\rangle, |z_{T0}\rangle\}$. See Fig. 1(c). $|z_S\rangle$ and $|z_{T0}\rangle$ are separated by the exchange energy (J), while $|x_-\rangle$ and $|x_+\rangle$ by $2(\mu_{vL} - \mu_{vR})|\mathbf{B_{normal}}|$. Here, $\mu_{vL}$, $\mu_{vR}$ = *valley magnetic moments* in the QDs, respectively.

This implementation is the analogue of the spin pair scheme[12-15]. It shares the distinctive advantages provided in the scheme, e.g., scalability and fault-tolerance, and the method developed in the scheme for initialization / readout / qugate operation[15] may be adapted here. The single qubit manipulation is discussed below.

**(Single qubit manipulation)** In the reduced space $\Gamma_v$, the qubit is described by the following effective Hamiltonan (in the basis of $\{|z_S\rangle, |z_{T0}\rangle\}$)

$$H_{eff} = (\mu_{vL} - \mu_{vR})B_{normal}\tau_x + \frac{J}{2}\tau_z. \tag{7}$$

The $\tau_x$ part generates a rotation $\check{R}_x(\theta_x = \Omega_x t_x)$ about the x-axis (of the corresponding Bloch sphere) when it is applied for the time $t_x$. Here, $\Omega_x$ is the "Larmor frequency", e.g.,

$$\Omega_x = 2(\mu_{vL} - \mu_{vR})B_{normal}/\hbar. \tag{8}$$

Similarly, the $\tau_z$ part generates a rotation $\check{R}_z(\theta_z = \Omega_z t_z)$ about the z-axis ($\Omega_z = J/\hbar$, and $t_z$ = corresponding time). $\check{R}_x$ and $\check{R}_z$, together, allow the qubit manipulation. See Fig. 2.

It is obvious from Eqn. (7) that $\mu_{vL} \neq \mu_{vR}$ is a required condition for the manipulation. There are various ways to generate this asymmetry. For example, a structural asymmetry between the QDs may induce a corresponding energy level difference and, hence, the required $\mu_v$ asymmetry according to Eqn. (2). Below, we discuss controllable, electrical means of tuning the asymmetry which are based on Eqns. (3) and (4).

DC Mode

A DC electric field is applied on one of the QDs, inducing $\delta\mu_v^{(DC)}$ to create $\mu_v$-asymmetry. An estimate using Eqns. (3) and (8), with the parameters $B_{normal}$=100mT, $k_{4x}=L^{-2}$ (L = QD size), $x_\varepsilon^{(DC)}/L= 0.2$, $\hbar w_0/\Delta=0.2$, $\Delta=0.14eV$,[24] gives $\Omega_x \sim O(ns^{-1})$, in the typical range currently envisioned in the QD approach.

AC mode

This mode is based on Eqn. (4). An AC field is superimposed on the DC field, and, for one half of the AC cycle, it induces the rotation $\check{R}_x(\theta_x^{(AC)})$, with

$$\theta_x^{(AC)} = \frac{2}{\hbar} \int_0^{\pi/w_{AC}} \delta\mu_v^{(AC)} B_{normal} dt \qquad (9)$$

$$= -\frac{3}{2} k_{4x} x_\varepsilon^{(DC)} x_\varepsilon^{(AC)} \frac{\mu_{v0} B_{normal}}{\hbar w_{AC}} \frac{\hbar w_0}{\Delta}.$$

For the other half cycle, the sign of $\theta_x^{(AC)}$ is flipped. Fig. 3 shows schematically how the qubit may be manipulated in the alternating sequence, $\check{R}_x(\theta_x^{(AC)}) \to \check{R}_z(\theta_z=\pi) \to \check{R}_x(-\theta_x^{(AC)}) \to \check{R}_z(\theta_z=\pi) \to \ldots$, into a target state. Under the condition $\Omega_z \ll w_{AC}$, the manipulation time is roughly

$$t_{AC} = O(1) \frac{t_z^{(0)}}{\theta_x^{(AC)}} \quad (t_z^{(0)} = \frac{\pi\hbar}{J}).$$

With $\hbar w_{AC}/J = 10$, $x_\varepsilon^{(AC)}/L = 0.2$, and the parameters used earlier in the estimation of $\Omega_x$ in DC mode, it gives $t_{AC} \sim O(10\text{ns})$.

**(Valley relaxation)** We estimate the valley relaxation rate $1/T_1$ in a QD, due to the intervalley scattering $K \leftrightarrow K'$.[30] The QD is assumed to be in the clean limit, where the structure is impurity-free, and the intervalley scattering occurs due to the QD potential. Let the valley splitting be $\delta E_{KK'} \sim 2\mu_{v0} B_{normal}$. At low temperatures, the relaxation process is mediated by acoustic phonons. A rough estimate yields[26]

$$\frac{1}{T_1} \sim O(1)(n_{Q_0} + \frac{1}{2} \pm \frac{1}{2}) \frac{\mu_{v0}^2}{e^2 \hbar^5 \rho_a c_s^8} \frac{1}{\left(1+\delta k^2 L^2\right)^2} \quad (for\ B_{normal} < \hbar w_0 / 2\mu_{v0}) \quad (10)$$

$$\left(\frac{V_0 D}{\hbar w_0 \Delta}\right)^2 \exp\left(-\frac{\hbar Q_0^2}{2m^* w_0}\right)(\mu_{v0} B_{normal})^6$$

$$(\hbar c_s Q_0 = 2\mu_{v0} B_{normal})$$

Here, $c_s$ = *sound velocity* ~ 2.1 x $10^4$ m/s, $\rho_a$ = *mass density*, L = *dot size*, $V_0$ = *QD potential depth*, D = *deformation potential constant* ~ 18eV, $\delta k$ = *wave vector difference* between K and K' points, and $n_{Q0}$ = *phonon occupation number* at wave vector $Q_0$. Using $\Delta$ ~ 0.14eV, L ~ 350Å, $V_0$ ~ 0.5$\Delta$ (for a bound state to exist), $\hbar w_0$ ~ 30meV, $B_{normal}$ ~ 100mT, and temperature = 10K, we obtain $T_1$ ~ O(ms), sufficiently long for qubit manipulation. We anticipate that experimental realization of the proposal here will lead to utilization of the valley d.o.f., in addition to the spin d.o.f., to encode quantum information, and unveil the intriguing prospect of valley-based quantum computing in carbon systems.

**Acknowledgment** – We thank the support of ROC National Science Council through the Contract No. NSC-99-2112-M-007-019.

**Competing financial interests** – The authors declare that there are no competing financial interests.

# Figure Captions

**Figure 1.**

(a) The DQD qubit structure. The QDs are electrostatically defined. Gate V' is used to tune the potential barrier and, hence, J. Gates $V_L$ and $V_R$ are applied for DC/AC $\mu_v$-tuning.

(b) The Zeeman-type interaction for spin and valley d.o.f.s, $H_Z = -g^* \sigma \mu_B |\mathbf{B}_{total}| + \tau_v \mu_v |\mathbf{B}_{normal}|$, splits the one-electron quadplet $|\tau_v = \pm 1, \sigma = \pm 1/2\rangle$. $g^*$ = *electron g-factor*, $\mu_B$ = *Bohr magneton*, $\mathbf{B}_{total} = \mathbf{B}_{plane} + \mathbf{B}_{normal}$, and $\sigma = \pm 1/2$ (denoting the spin states quantized along $\mathbf{B}_{total}$). For $g^* \mu_B |\sigma \mathbf{B}_{total}| > \mu_v |\mathbf{B}_{normal}|$, the splitting leaves $|\tau_v = \pm 1, \sigma = 1/2\rangle$ as the lower doublet to form singlet / triplet states. (The index $\sigma = 1/2$ is dropped in the text.)

(c) $|z_{T\pm}\rangle$ are split away from $\{|z_S\rangle, |z_{T0}\rangle\}$, by $\pm (\mu_{vL} + \mu_{vR})|\mathbf{B}_{normal}|$, respectively. $|z_S\rangle$ and $|z_{T0}\rangle$ are split in energy by J, while $|x_-\rangle$ and $|x_+\rangle$ by $2(\mu_{vL} - \mu_{vR})|\mathbf{B}_{normal}|$.

**Figure 2.**

The time evolution of a qubit state, as governed by $H_{eff}$, consists of a rotation $\check{R}_x(\theta_x)$ about the x-axis of the Bloch sphere, and a rotation $\check{R}_z(\theta_z)$ about the z-axis. $\theta_x$ and $\theta_z$ are the respective angles of rotation.

**Figure 3.**

In the AC mode, the initial qubit state, e.g., $|z_S\rangle$, may be manipulated in the alternating sequence, $\check{R}_x(\theta_x^{(AC)}) \rightarrow \check{R}_z(\theta_z=\pi) \rightarrow \check{R}_x(-\theta_x^{(AC)}) \rightarrow \check{R}_z(\theta_z=\pi) \rightarrow \ldots \check{R}_z(\theta_z^{(target)}+\pi/2)$, into a target state ($\theta_z^{(target)}$ = *target state longitude*).

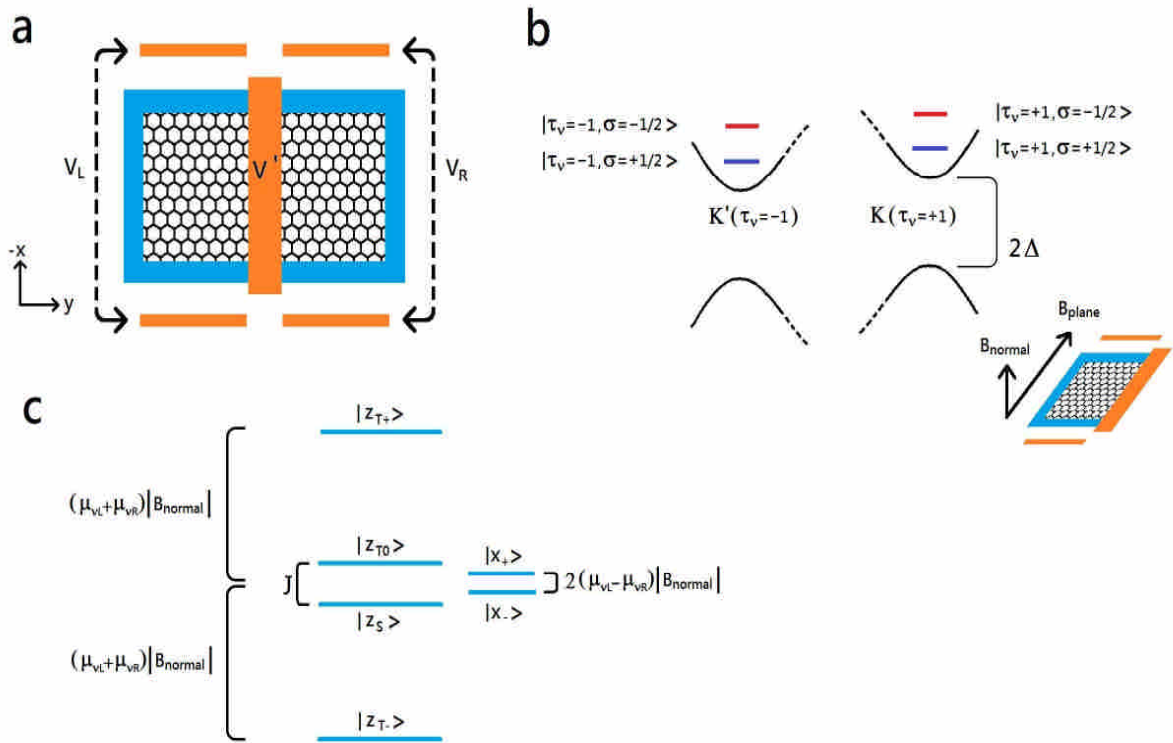

Figure 1

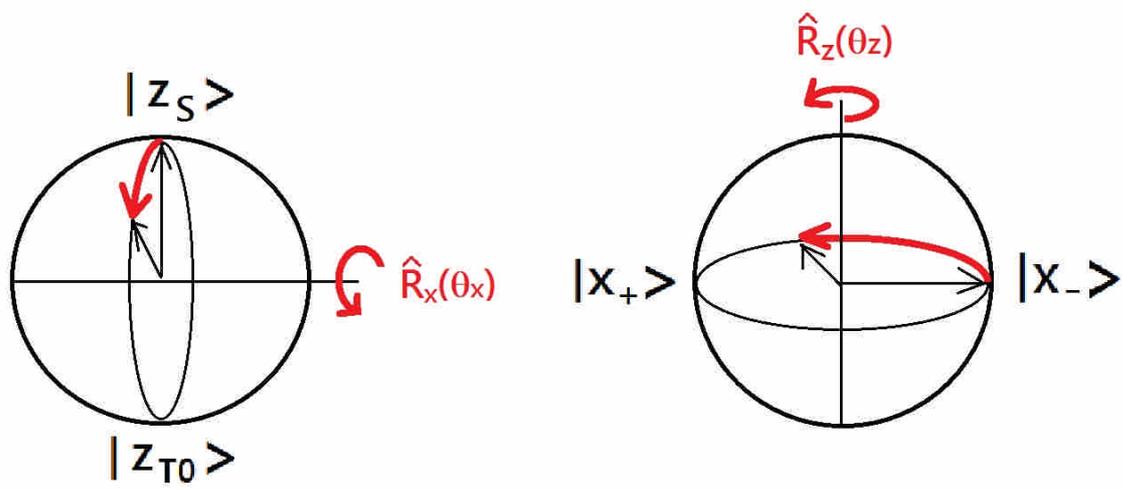

Figure 2

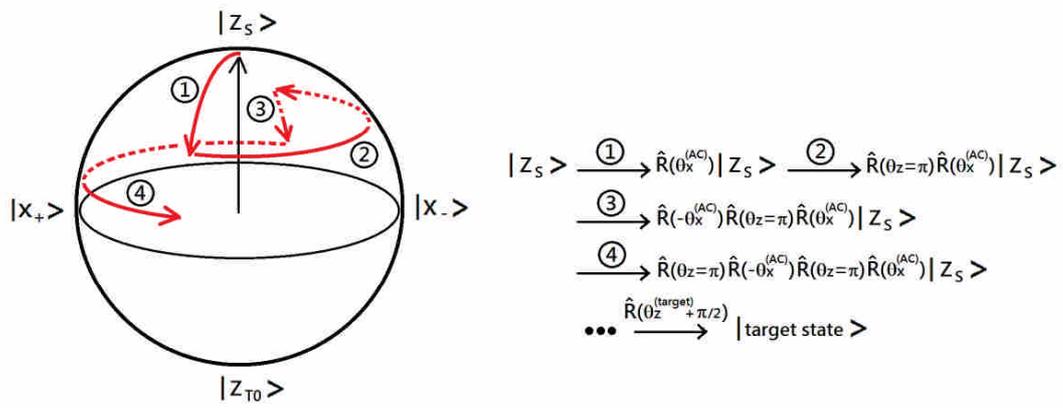

Figure 3